# Trion quantum coherence in site-controlled pyramidal InGaAs quantum dots

R. A. Barcan[1,2,†], I. Samaras[1,†], K. Barr[1], G. Juska[3], E. Pelucchi[3], K. G. Lagoudakis[1,*]

[1] *Department of Physics, University of Strathclyde, Glasgow G4 0NG, United Kingdom*
[2] *Sussex Centre for Quantum Technologies, University of Sussex, Brighton BN1 9QH, United Kingdom*
[3] *Tyndall National Institute, University College Cork, Cork T12 R5CP, Ireland*

[†] These authors contributed equally to this work
* k.lagoudakis@strath.ac.uk

## Abstract

Deterministically positioned pyramidal InGaAs quantum dots (QDs) exhibit exceptional quantum properties, making them highly promising candidates for scalable on-chip quantum information processing. In this work, we investigate the coherent dynamics of positively charged excitons under the influence of strong magnetic fields in the Faraday configuration. Pyramidal quantum dots exhibit a fourfold splitting of the charged excitons even in the Faraday configuration, giving rise to an optically addressable double-Lambda ($\Lambda$) system akin to self-assembled quantum dots in oblique magnetic fields. Here, we investigate ultrafast complete coherent control of the trion to ground state transition utilizing advanced optical resonant excitation techniques and we observe quantum coherence over timescales that are similar to other prominent quantum dot platforms. These results pave the way towards establishing site-controlled pyramidal InGaAs QDs as scalable platforms for quantum information processing, expanding the reach of coherent control to new quantum systems.

## Introduction

Epitaxially grown semiconductor quantum dots (QDs) represent a promising platform for the realization of on-chip quantum hardware. However, achieving large scale integration remains a significant challenge, primarily due to the difficulty to grow high quality QDs at on-demand predefined locations, while simultaneously ensuring high optical quality, spectral homogeneity, and long spin coherence times suitable for spin qubit arrays.

This requirement has fueled extensive research efforts toward developing site-controlled quantum dot arrays, enabling deterministic positioning while maintaining all the advantageous properties of self-assembled QDs and facilitating seamless integration into scalable quantum architectures[1-10].

Among existing scalable QD platforms, site-controlled QDs grown using Metal-Organic Vapor Phase Epitaxy (MOVPE) within (111)-oriented GaAs pyramidal recesses stand out as particularly promising candidates. These QDs offer numerous advantages, such as high optical quality with narrow emission spectra, good uniformity of the emission energy[11], single and polarization-entangled photon emission[12,13], along with spatial control, and wavelength tunability which is influenced by the composition and growth conditions[10]. Notably, site-controlled pyramidal QDs also demonstrate intrinsic charge retention capabilities, a feature that holds significant potential for enabling the generation and coherent manipulation of spin qubits under the influence of external magnetic fields, which lifts the degeneracy of the hole ground state and the trion, producing an optically addressable double-$\Lambda$ level structure that is characteristic of QDs possessing $C_{3v}$ symmetry, as realised in (111)-oriented heterostructures[14,15].

The coherence properties of the optically addressable qubits are typically characterized by measuring the dephasing time $T_2^*$. This measurement, often relies on multi-pulse excitation with precise amplitude and phase control of the optical fields. Such interferometric techniques are widely utilized to assess dephasing dynamics across a diverse range of optically addressable quantum emitters, including quantum dots, atomic systems, and defect centers in solid-state materials[15-30].

In this work, we set out to investigate the coherence properties of positively charged excitons (trions) in this versatile quantum dot platform. By applying a magnetic field in the Faraday configuration, we induce Zeeman splitting of the ground hole and excited trion states, giving rise to an optically addressable double-$\Lambda$ structure, typical for quantum dots grown on (111)-oriented substrates[14]. Taking advantage of this configuration, we use a combination of weak continuous-wave (CW) above-band (Ab. Bd.) excitation and resonant driving with ultrafast optical pulses, in combination with interferometric coherent manipulation techniques to coherently control and characterize the dephasing properties of the trion to ground-hole-spin state qubit.

## Results

### Rabi Oscillations

The ability to manipulate the qubit state is one of the vital foundations of quantum information processing. Qubit rotations around one specific axis in the Bloch sphere are conveniently manifested with the observation of Rabi oscillations. In this experiment, we aim to demonstrate Rabi rotations of the QD's positively charged trion to hole ground state system using ultrafast optical pulses, resonant with the $|\Downarrow\rangle_\vartheta \rightarrow |\Uparrow\Downarrow\downarrow\rangle_\vartheta$ transition. Observation of the coherent rotations is performed taking advantage of the Λ-system structure, detecting photons from the $|\Uparrow\Downarrow\downarrow\rangle_\vartheta \rightarrow |\Uparrow\rangle_\vartheta$ transition via radiative decay, which is spectrally separated from the driven transition, as shown in **(Fig. 1(a))**. The resonant pulses, generated by a Ti:Sapphire laser, have a fixed duration ($\sim$3 ps), while the pulse area is varied to control the rotation angle θ about the x-axis on the Bloch sphere. By adjusting the pulse area which is proportional to the rotation angle and depends on the square root of the pulse power, coherent rotations should be observed between the ground state $|\Downarrow\rangle_\vartheta$ and the excited trion state $|\Uparrow\Downarrow\downarrow\rangle_\vartheta$.

Since the trion state has a finite lifetime of approximately 1 ns, any population on the excited state eventually decays via spontaneous emission to either hole ground state, producing measurable photons. Weak Ab. Bd. excitation randomizes the ground state populations enabling the repetition of the experiment. To ensure that the desired transition is being driven without excitation of higher energy states, the resonant laser is carefully tuned to overlap with the lowest energy transition while eliminating the high energy pulse tail with a pulse shaper **(Fig. 1(a))**. Plotting the detected photon counts as a function of the resonant pulse power, reveal distinct Rabi oscillations as the pulse area is increased from 0 to approximately $4\pi$ **(Fig. 1(b))**. At the first maximum, corresponding to a $\pi$-pulse, the photon count peaks, reflecting transfer of the population to the excited state ($|\Downarrow\rangle_\vartheta \rightarrow |\Uparrow\Downarrow\downarrow\rangle_\vartheta$). This corresponds to a half rotation ($\theta = \pi$) on the Bloch sphere. In contrast, at the first minimum, corresponding to a $2\pi$-pulse, the photon count reaches a minimum, as the system completed a full Rabi cycle ($\theta = 2\pi$ on the Bloch sphere) and returned to the ground state within the duration of the pulse, resulting in minimum population in the excited state. Further increasing resonant pulse area reveals oscillations up to $4\pi$, corresponding to two complete cycles of population transfer between the ground state and excited state. In an ideal system a resonant $2\pi$ pulse would return the Bloch vector exactly to the ground state, yielding zero fluorescence. However, in our system, the coupling to longitudinal acoustic phonons introduces pulse-area-dependent pure dephasing that shortens the Bloch vector. Consequently, the fluorescence minimum at $2\pi$ remains finite and the contrast of the Rabi oscillations decays, in agreement with the microscopic theory of Förstner et al.[31].

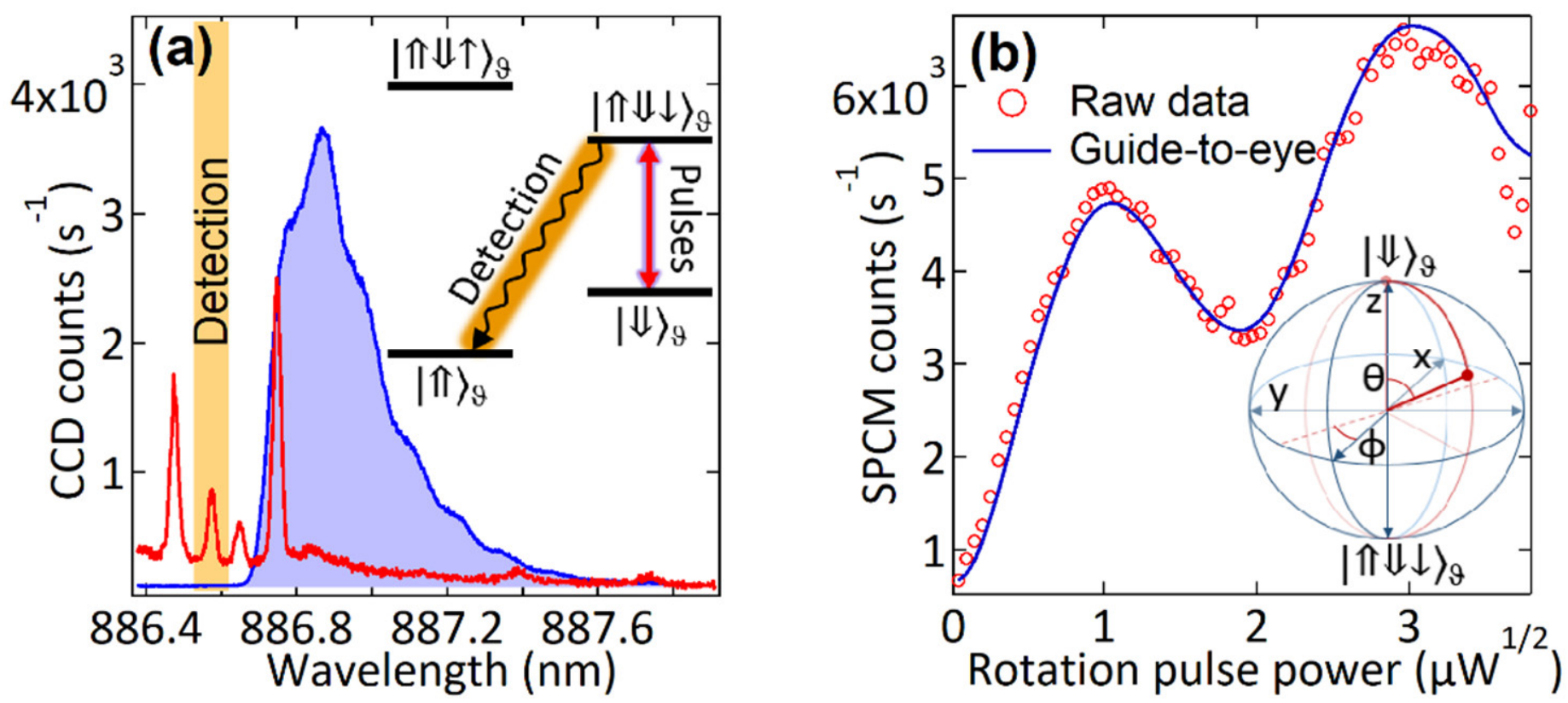

**Figure 1**: **(a)** Photoluminescence spectra showing the all-optical driving and detection scheme of the coherent control experiment. Transition $|\Downarrow\rangle_\vartheta \rightarrow |\Uparrow\Downarrow\downarrow\rangle_\vartheta$ of our qubit is resonantly addressed with spectrally filtered pulses from a Ti:Sapphire Laser (blue pulse), while the emission from the transition$|\Uparrow\Downarrow\downarrow\rangle_\vartheta \rightarrow |\Uparrow\rangle_\vartheta$ is detected through the spectrometer by a single-photon counting module (SPCM) for measuring photon count rates (yellow area). **(b)** Plot of Rabi oscillations as a function of the pulse power. The oscillations demonstrate the ability to all-optically control the trion to ground state qubit by varying the pulse area. The inset depicts the corresponding rotations of the Bloch vector from the ground state to the excited state along the red trajectory defined by the Bloch vector's rotation.

An observed slight monotonous increase of the intensity is possibly related to the deviations from a transform-limited Gaussian pulse introduced during the pulse shaping or incoherent contributions relevant at higher driving powers[32]. These observations provide clear evidence of coherent control and demonstrate the system's ability to maintain coherence while performing quantum operations.

**Ramsey Interference**

Having identified the characteristic pulse areas corresponding to $\pi$, $2\pi$, and higher-order rotations, we shift our focus to investigating the coherence properties of the quantum dot by measuring Ramsey interference which gives insights to the trion decoherence time $T_2^*$. Instead of applying a single pulse, two identical pulses, each with a pulse area corresponding to $\theta = \pi/2$, are employed and separated by a controlled time interpulse delay $\tau$. The first $\pi/2$-pulse creates a coherent superposition between the ground hole state $|\Downarrow\rangle_\vartheta$ and the excited trion state $|\Uparrow\Downarrow\downarrow\rangle_\vartheta$, placing the system in a state where the quantum states maintain well-defined phase relationship with equal probability amplitudes. During the interval between the two pulses, the system evolves under its internal perturbations, precessing at the optical frequency $\omega$ of the transition $|\Uparrow\Downarrow\downarrow\rangle_\vartheta \rightarrow |\Downarrow\rangle_\vartheta$ and therefore this system is best described using the rotating frame approximation. Therefore, in the rotating frame the second $\pi/2$ pulse will rotate the Bloch vector by $\pi/2$ around an axis which is at an angle $\phi$ with respect to the x axis[28,29]. Each delay corresponds to a specific angle $\phi$ of the rotation axis around which the second pulse rotates the Bloch vector, with $\phi$ directly proportional to the interpulse delay $\tau$. To control and vary the interpulse delay $\tau = \tau_c + \tau_f$, a Mach-Zehnder interferometer is employed.

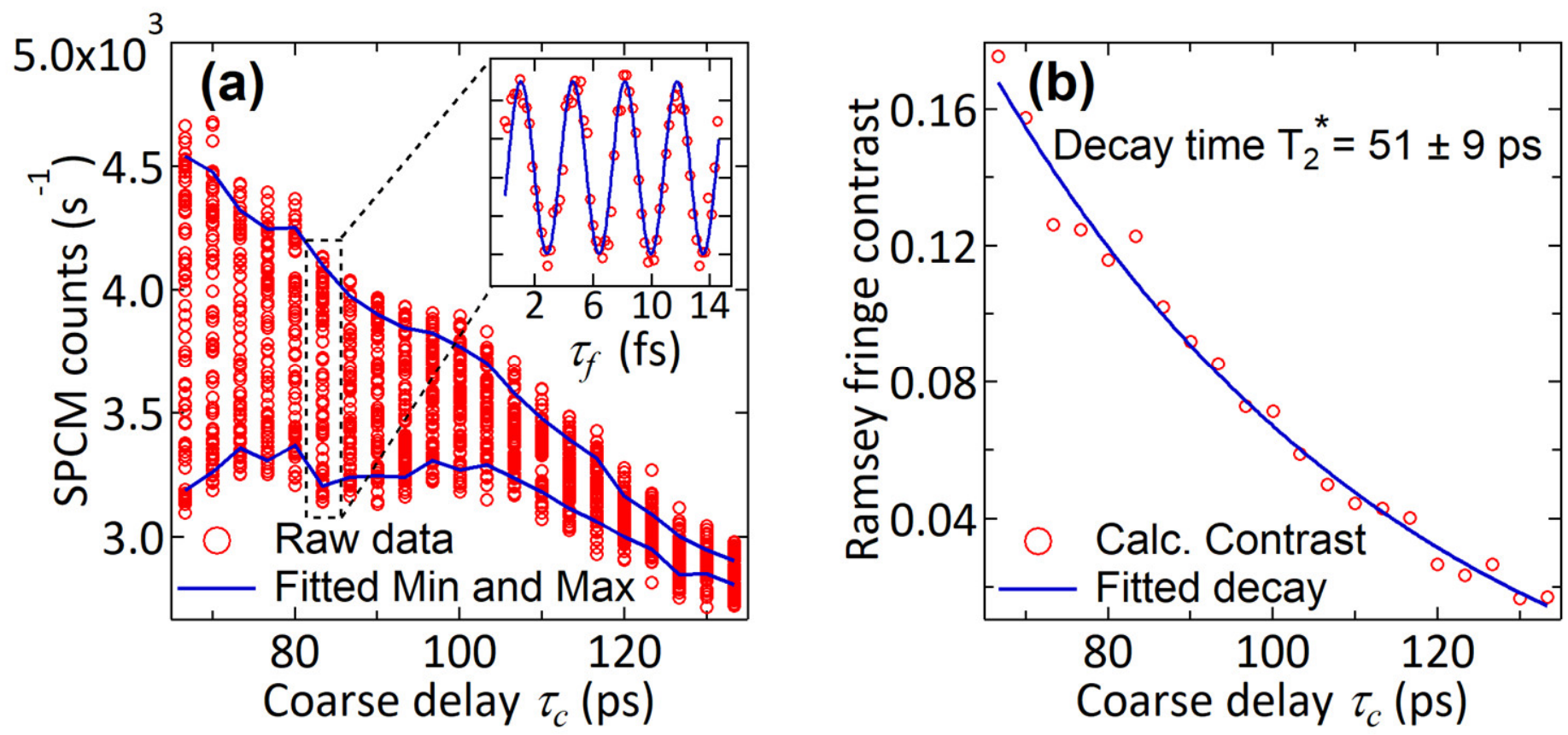

**Figure 2**: **(a)** Ramsey fringes as a function of coarse delay. The raw data points are depicted by red circles, while the blue lines represent the maximum and minimum intensity values of the fitted fringes. The inset shows an expanded view of the Ramsey fringes as a function of the fine delay $\tau_f$, taken at coarse interpulse delay $\tau_c = 83.4$ ps, clearly highlighting the oscillatory behavior due to quantum interference. **(b)** Decay of the Ramsey fringe contrast plotted against coarse delay. The red circles correspond to the calculated (calc.) contrast obtained from the fitted maximum and minimum intensities of the Ramsey fringes at each coarse delay in panel (a). The blue line is an exponential decay fit to the data.

The coarse delay $\tau_c$ is introduced by adjusting a motorized delay stage in one arm of the interferometer, while the fine delay $\tau_f$ is precisely controlled using a piezoelectric stage to introduce an additional minute delay in the other arm. When the delay satisfies $\phi = 2n\pi$, the second $\pi/2$-pulse rotates the Bloch vector into the excited state $|\Uparrow\Downarrow\downarrow\rangle_\vartheta$, maximizing the detected photon counts. Conversely, when $\phi = (2n + 1)\pi$, the second pulse returns the system to the ground state $|\Downarrow\rangle_\vartheta$, resulting in minimal photon counts from the detected $|\Uparrow\Downarrow\downarrow\rangle_\vartheta \rightarrow |\Uparrow\rangle_\vartheta$ transition. By varying the interpulse delay, first by adjusting the coarse delay and then fine-tuning with the piezoelectric actuator, oscillations in the photon counts as a function of fine delay are observed, manifesting as Ramsey interference fringes **(Fig. 2(a))**. The experimental setup enables systematic variation of the coarse delay starting at $66.7$ ps, with $3.33$ ps increments, while fine adjustments are achieved using a piezoelectric actuator that introduces an additional optical path difference of $5.33\ \mu\mathrm{m}$, translating to a time delay of approximately $14$ fs. The initial coarse delay is set to eliminate any optical interference between the pulses, ensuring that the observed oscillations in the detected photon count rate are attributable to quantum mechanical effects of the system rather than classical interference. By analyzing the Ramsey fringes and plotting the fringe contrast values against the coarse delay **(Fig. 2(b))**, the decoherence time ($T_2^*$) of the trion-ground state qubit is evaluated. We use the standard definition for contrast, using the max and min fitted values (blue lines) of the Ramsey Fringes in **(Fig. 2(a))**. Fitting the decay of the fringe contrast with a single exponential function yields a coherence time of $T_2^* = 51 \pm 9$ ps. This value is consistent with previously measured coherence times in other QD systems[28].

**Complete Coherent Control**

The complete coherent control dynamics experiment aims to demonstrate universal single qubit gate operations by enabling access to the entire Bloch sphere[28,29]. This is accomplished using an experimental configuration similar to that of the Ramsey interference setup, with a key distinction: the pulse areas are no longer fixed at $\pi/2$ but are tunable to arbitrary rotation angles. It can be shown that two consecutive rotations by an angle $\theta$, with the first rotation taking place around the x-axis and the second rotation taking place around an axis that is rotated by $\phi$ with respect to the x-axis, are sufficient to reach any point on the Bloch sphere, thus enabling universal qubit control. In the experiment, both the pulse area (proportional to the rotation angle $\theta$) and the fine interpulse delay $\tau_f$ between the pulses (proportional to the phase angle $\phi$) are systematically varied. The resulting data are represented in a two-dimensional colour plot **(Fig. 3(a))**, with the horizontal axis corresponding to pulse power and the vertical axis to the fine delay, thereby mapping the full coherent evolution of the qubit state. The detected photon counts from the $|\Uparrow\Downarrow\downarrow\rangle_\vartheta \rightarrow |\Uparrow\rangle_\vartheta$ transition are presented as the colourscale. The coarse delay is fixed at $66$ ps, sufficiently long to prevent optical interference between the pulses. The two rotation operations are achieved by setting the pulse power, followed by scanning the fine interpulse delay $\tau_f$ over approximately $12$ fs, while recording the detected photon counts for each delay point. This interference experiment is performed across all accessible pulse powers in the range of 0 to $2.5\ \mu\mathrm{W}^{1/2}$ and the resulting photon counts were compiled into a photon-count map **(Fig. 3(a))**. As this is an interferometric measurement requiring an identical range of phase delays for all pulse powers, it is crucial that any phase drift that occurs over the duration of this experiment ($\sim 4$ h) is recorded and corrected. To quantify and correct unwanted phase drifts during this lengthy experiment, a frequency-stabilized CW

HeNe laser is introduced into the Mach–Zehnder interferometer in a configuration similar to Okada *et al.*[24] , but here counter propagating with respect to the rotation pulses. As the HeNe beam is aligned to counter-propagate through the same interferometric path, both the rotation pulses and the HeNe laser experience identical path-length variations and perturbations during the experiment. This enables for continuous access to the relative phase between the pulses throughout the entirety of the duration of the complete coherent control experiment. The red line in **(Fig. 3(b))** shows the unwrapped phase drift of the HeNe laser for the duration of the experiment. Scaling this for the wavelength of the emission of the detected photons and transforming it to a time delay, allows to effectively correct the unwanted phase drift and access the complete coherent control behavior as fully attributable to quantum mechanical effects of the system free from external experimental artifacts. The initial HeNe phase clearly illustrates the presence of accumulated drifts, manifesting as a gradual phase distortion over time. In contrast, after applying the correction, the final unwrapped phase shown in blue in **(Fig. 3(b))**, exhibits a stable and well-structured evolution, with minor noise being present, demonstrating the effectiveness of the correction process. Here, we note that the phase drift correction process was applied post-acquisition and not during the experiment. From the unwrapped, zero-centered HeNe phase, we find a residual RMS phase error of $\sigma_{ph.} = 0.18$ rad, indicated by the pink solid band behind the corrected phase trace in **(Fig. 3(b))**. The stark contrast between the initial and corrected phase drift highlights the necessity of precise phase stabilization techniques to mitigate unwanted drifts, preserve coherence, and ensure reliable relationships between optical pulses.

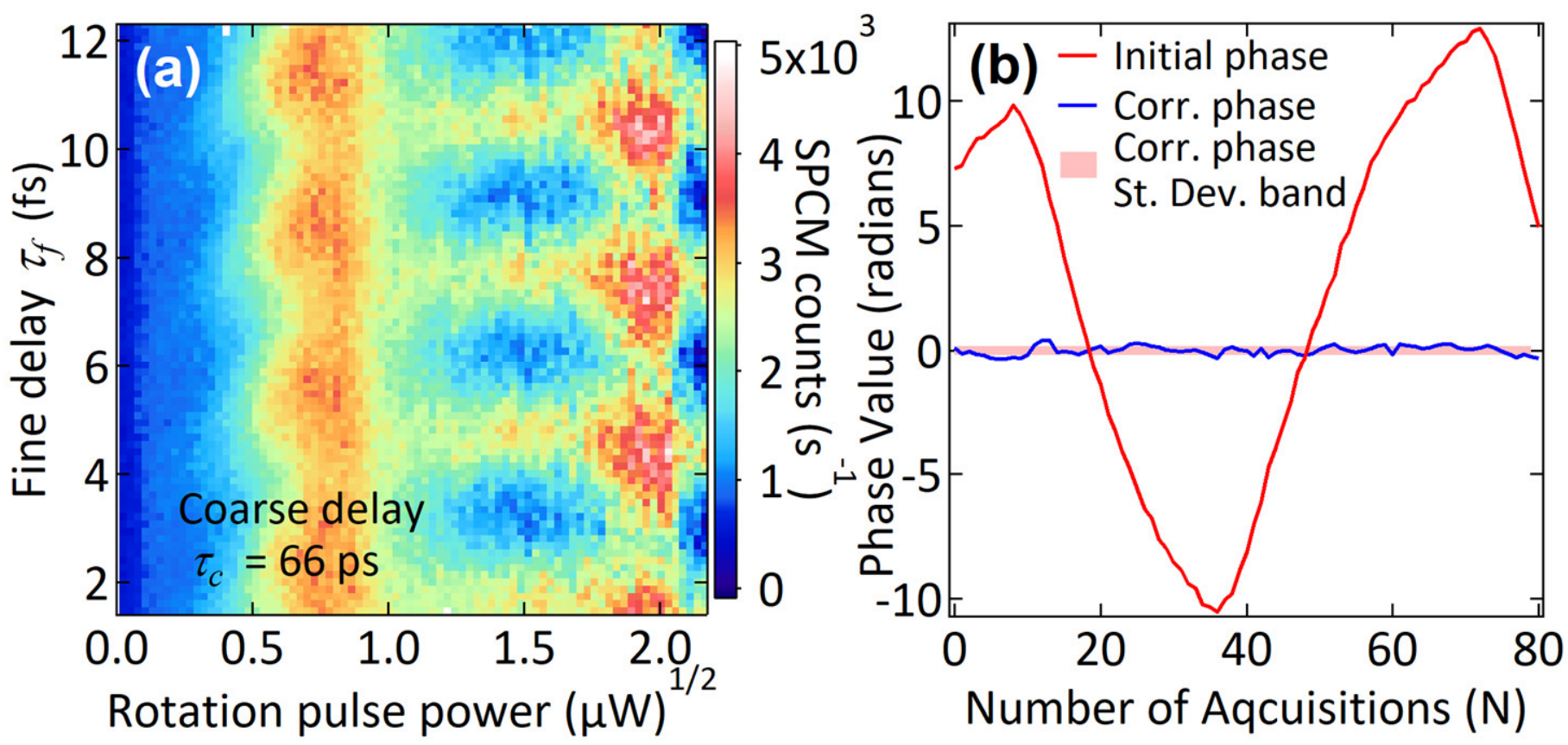

**Figure 3**: **(a)** Map of SPCM counts obtained from the SU(2) experiment illustrating the variation in observed counts as a function of fine interpulse delay $\tau_f$ (vertical axis) and resonant pulse power (horizontal axis). The multi-lobed pattern signifies coherent control of the trion to ground state system, demonstrating the full SU(2) dynamics. The interferometric drift of the data during the long acquisition has been corrected according to the recorded phase shift of the reference HeNe Laser. **(b)** Recorded phase of the HeNe laser during the experiment. The red trace represents the initial, uncorrected data, while the blue trace corresponds to the corrected (corr.) phase after compensating for the drift. The pink band corresponds to the standard deviation of the corrected phase values. The width of the band is here $2\sigma_{ph.}$.

Such refined phase control, is crucial for enabling deterministic quantum gate operations in resonantly driven systems and ultimately improves the stability of SU(2) dynamics in quantum information processing. The corrected complete coherent control data reveal the characteristic multilobed structure of the SU(2)[28,29], demonstrating successful control over the quantum states. Ultimately, these results, demonstrate the ability to fully coherently control the trion qubit states in these QDs.

## Discussion

We demonstrate that site-controlled InGaAs pyramidal quantum dots (QDs) are highly promising scalable quantum emitters for quantum hardware applications. Through the use of ultrafast resonant pulses and Ab. Bd. excitation, we show coherent control of the positively charged excitonic complex. Our experiments show that we can perform ultrafast Rabi rotations of the trion to hole-ground-state qubit, while for dual pulse driving with variable delays we observe Ramsey interference, with a dephasing time that is similar to other quantum dot platforms. Combining dual pulse driving with variable amplitudes and delays, we demonstrate full access to the Bloch sphere, confirming complete coherent control of the trion to spin-ground-state qubit states. These results lay the groundwork towards utilizing the long lived ground state spins as optically addressable spin qubits, with direct applications in geometric phase gates[33], establishing site-controlled pyramidal QDs as promising candidates for on-chip quantum information processing. Our findings contribute to the foundations for developing robust quantum technologies, providing essential insights into the manipulation and control of qubit states in promising solid-state systems.

## Methods

The investigated sample consists of arrays of $In_{\mathrm{x}}Ga_{1-\mathrm{x}}As$ quantum dots (QDs) with a nominal thickness of $0.85\ \mathrm{nm}$ and Indium concentration of $x = 0.25$. Because of alloy segregation however, the effective quantum dot thickness is $\sim 3.4\ \mathrm{nm}$ and indium concentration is $x \approx 0.29$[34]. These quantum dots are grown on a (111)B-oriented GaAs substrate pre-patterned with tetrahedral pyramidal recesses as depicted in the sample rendering in **Fig. 4(a)**, where we have overlayed a white light microscopy image on top of it for enhanced clarity. This site-controlled growth technique utilizes advanced lithography to achieve precise positioning and uniform sizing of the QDs, resulting in high spectral uniformity and purity-attributes that are critical for high-performance quantum emitters. To enable magnetic cryomicroscopy studies of the quantum dots, the sample is housed within a magnetic cryostat (Oxford Microstat-MO) integrated with a helium cryorecirculator (ColdEdge Stinger) for closed-cycle operation [35]. The experimental setup provides a continuous supply of ultra-cold helium, ensuring stable cryogenic conditions required for the operation of both the superconducting magnet and the sample. The sample temperature is maintained at $6.8 \pm 0.1\ \mathrm{K}$. Prior to selecting a specific quantum dot (QD) for detailed investigation, the sample is characterized at zero magnetic field using a custom-designed scanning confocal microscopy system. This apparatus allows for simultaneous scanning of both the excitation and collection spots while recording photoluminescence (PL) photon counts above 800 nm. The system employs a long working distance microscope objective lens (approximately 12 mm) with a numerical aperture (N.A.) of 0.5, in conjunction with a dual-axis galvanometric scanner to facilitate precise spatial mapping. For the generation of the scanning confocal microscopy map (**Fig. 4(b)**), Ab. Bd. excitation is provided by a 780 nm laser diode exciting the sample at a power of about 200 nW, which is close to the typical saturation power of QDs in this sample. As seen in **Fig. 4(b)** the pitch of the pyramidal recesses containing individual QDs is 7.5 μm, well above the diffraction limit of our setup, enabling individual dot addressability. PL from the QDs is spectrally filtered using a custom-designed double Czerny-Turner spectrometer with intermediate slits. This system, consisting of two *1* m - long stages in series, achieves a high spectral resolution of approximately 8 μeV.

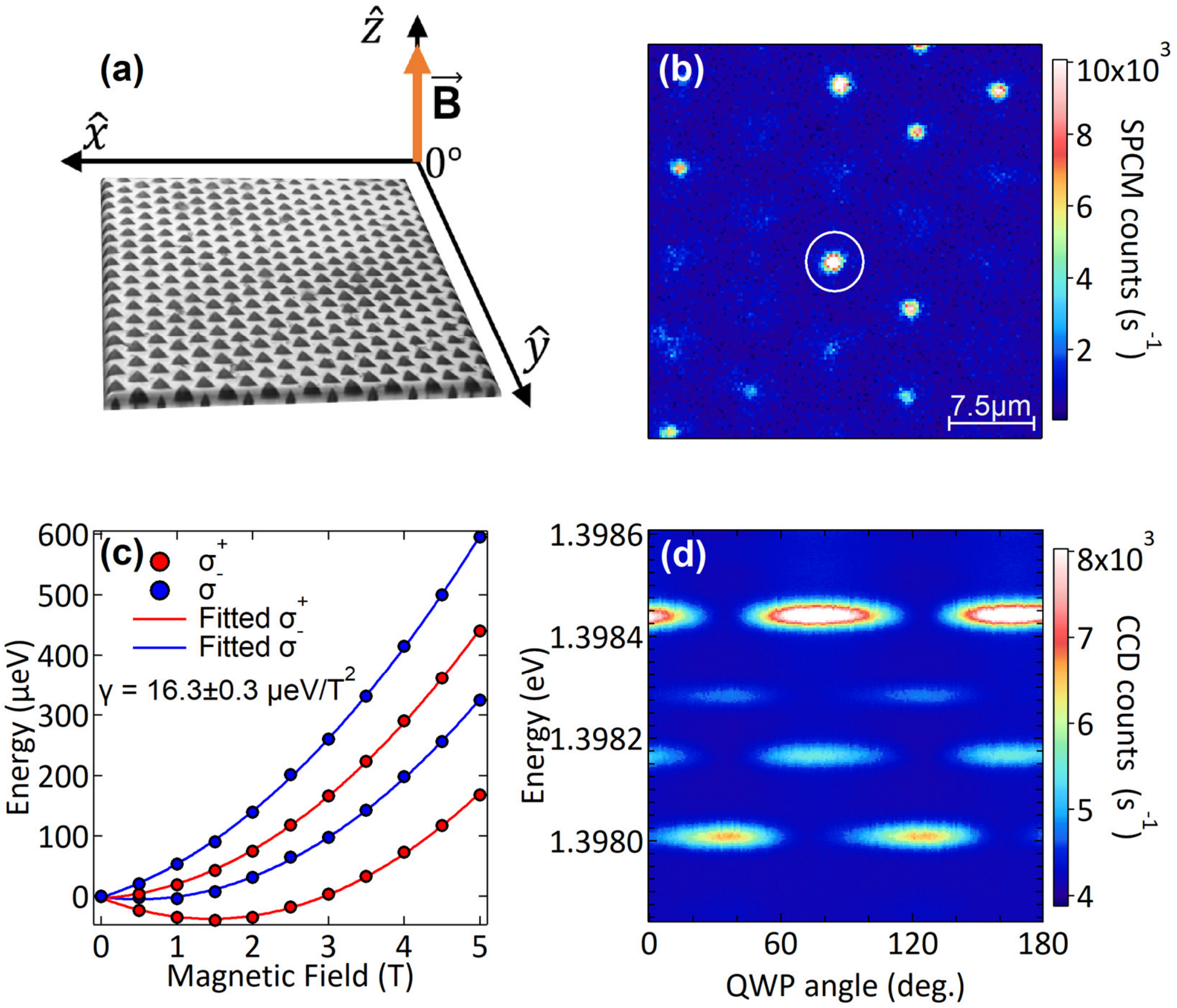

**Figure 4**: **(a)** Diagram showing the orientation of the magnetic field with respect to the sample plane. The rendered structure has been overlayed with a white-light microscopy image of the actual sample surface for illustrative clarity. **(b)** Two-dimensional PL map of the scanned QD array, with emission intensity represented on a colourscale in counts per second on the SPCM. The orientation of the PL-map matches the white-light microscopy image in (a). **(c)** Gradual splitting of the emission spectra as the magnetic field strength increases, exhibiting a non-linear relationship between energy and magnetic field, characteristic of the Zeeman effect and the diamagnetic shift. **(d)** Polarization analysis of the four resulting transitions at 5 T. The colourscale represents the relative intensity of each transition.

The spectrometer utilizes $1800\ \text{lines/mm}$ diffraction gratings, resulting in an overall optical efficiency of around 50%. The filtered PL signal is directed either to a charge-coupled device (CCD) array for capturing wavelength spectra or to single-photon counting module (SPCM) for measuring photon count rates at specific energies. Narrowing the intermediate slits of the spectrometer, allows for selective detection of the inner transition with the highest energy. Photon counts are continuously optimized during measurements to compensate for potential losses caused by the magnetic field's influence on critical optical components such as the microscope objective lens. Both the pulsed and the Ab. Bd. laser powers are finely controlled using motorized linear polarizers, allowing precise adjustment of the excitation intensity.

With the superconducting magnet energized, the cryostat is capable of generating a homogenous magnetic field of up to $5\ \text{T}$, parallel to the growth axis of the sample (Faraday configuration). This magnetic field acts on [111]-oriented quantum dots (QDs) in a manner analogous to an oblique-field configuration in conventional [001] QD structures. In [001] QDs, symmetry prohibits heavy-hole (HH) mixing under a purely longitudinal field; by contrast, the $C_{3v}$ symmetry, triangular confinement, and absence of a mirror plane in [111] dots introduce an off-diagonal Zeeman term that couples the HH basis states through an effective g-factor. This interaction produces a four level double-$\Lambda$ system[14,15,36]. The resulting hole eigenstates are coherent superpositions $C_1|\Uparrow\rangle + C_2|\Downarrow\rangle$ and $C_1|\Uparrow\rangle - C_2|\Downarrow\rangle$ where the coefficients $C_1$ and $C_2$ are set by the ratio of the diagonal to off-diagonal g-factors[36]. Because the mixing stems from the dot's intrinsic shape anisotropy rather than an external perturbation, it is fundamentally distinct from the behaviour of [001] systems. In the latter, HH mixing can be induced by tilting the applied magnetic field. The transverse magnetic field component mixes the HH states, recreating an equivalent four-level double-$\Lambda$ structure[37] with similar properties to the pyramidal QDs. Based on this similarity and to simplify the notation, we use the symbols $|\Uparrow\rangle_\vartheta$ and $|\Downarrow\rangle_\vartheta$ to depict the coherent superpositions of the holes in the ground states of the pyramidal QDs[15].

**(Fig. 4(c))** shows the evolution of the splitting for two cross-circular polarizations as a function of the magnetic field strength. The overall behavior is a result of the linear-in-field Zeeman splitting coupled to the quadratic-in-field diamagnetic shift. After fitting the four lines we estimate the diamagnetic shift factor as $\gamma = 16\ \mu\text{eV/T}^2$ and the effective g-factors for the electron and hole for this quantum dot as $g_e^{eff} = 0.54$ and $g_h^{eff} = 0.94$. At field, we use rotating quarter-wave polarimetry to further characterize the polarization properties of the four transitions which provides additional understanding of the selection rules governing the double-$\Lambda$ system **(Fig. 4(d))**. We find that the two $\Lambda$-systems have their transitions cross circularly polarized with a relatively high degree of circular polarization for all four transitions (minimum 93%).

Ultrafast all-optical coherent control of the QD energy states is achieved using a pulsed Ti:Sapphire laser (Spectra Physics, Tsunami), which produces $3\ \text{ps}$ pulses at $80.2\ \text{MHz}$ repetition rate. The pulses are sent through a pulse shaper to ensure that the output wavelength matches the lowest energy transition, while the high energy tail of the pulse is eliminated to ensure minimal overlap with the next in energy transition.

For the coherent control experiments we additionally use a weak CW Ab. Bd. excitation of the quantum dot, with $P_{Ab.Bd.} = 10\ \text{nW}$, corresponding to 4% of the saturation power for this QD, which effectively randomizes the ground state populations, preventing the depletion of any particular ground state because of spin initialization[15].

To isolate the Rabi oscillations and eliminate background effects, an additional control experiment is conducted without the Ab. Bd. randomization laser, thereby allowing initialization of the system in the $|\Uparrow\rangle_{\vartheta}$ state and leading to the collection of background counts. By subtracting the background counts from the original data with both lasers on, the emission intensity solely due to the resonant pulses is retrieved.

## Funding

R.A.B. acknowledges financial support from the EPSRC Doctoral Training Program under grant no. EP/Y011864/1 and the Agency for Student Loans and Scholarships (https://roburse.ro) under the scholarship H.G. no. 118/2023. I.S. acknowledges financial support from the EPSRC Doctoral Training Partnership under grant no. EP/W524670/1. K.B. acknowledges financial support from the EPSRC Doctoral Training Program under grant no. EP/R513349/1. G.J. and E.P. acknowledge funding from Research Ireland, formerly Science Foundation Ireland, under Grants Nos. 22/FFP-P/11530, 22/FFP-A/10930, 15/IA/286

## Contributions

R.A. Barcan, I. Samaras, and K.G. Lagoudakis acquired and analyzed the data, and drafted the initial manuscript. K. Barr carried out the preliminary characterization of the samples. G. Juska and E. Pelucchi grew and optimized the samples that underpin the experiments. All authors engaged in frequent discussions of the results, contributed to revising the manuscript for intellectual content, and approved the final version for submission.

## Data availability

Data sets generated during the current study are available from the corresponding author on reasonable request.